\documentclass[onecolumn, 12pt, draftcls]{IEEEtran}
\usepackage[margin=1.02in]{geometry}
\usepackage{epsfig}
\usepackage{graphicx}
\usepackage{amsmath}
\usepackage{amssymb}
\usepackage{verbatim}
\usepackage{mathrsfs}
\usepackage{lipsum}
\usepackage{graphicx}
\usepackage{subcaption}
\usepackage{amsmath,amssymb}
\usepackage{amssymb}%
\usepackage{mathrsfs}%
\usepackage{blindtext}
\usepackage{xcolor}
\usepackage{colortbl}
\usepackage{balance}
\usepackage{booktabs}
\usepackage{cite}
\IEEEoverridecommandlockouts
\def\BibTeX{{\rm B\kern-.05em{\sc i\kern-.025em b}\kern-.08em
    T\kern-.1667em\lower.7ex\hbox{E}\kern-.125emX}}
\begin{document}
\title{Deep Learning based End-to-End Wireless Communication Systems with Conditional GAN as Unknown Channel}
\author{\IEEEauthorblockN{Hao Ye, Le Liang, Geoffrey Ye Li, and Biing-Hwang Fred Juang
}

\thanks{This work was supported in part by a research
gift from Intel Corporation and the National Science Foundation under Grants
1815637 and 1731017.

H. Ye, G. Y. Li, and B-H. F. Juang are with the School of Electrical and Computer
Engineering, Georgia Institute of Technology, Atlanta, GA 30332 USA (email: yehao@gatech.edu; liye@ece.gatech.edu; juang@ece.gatech.edu)

L. Liang is with Intel Labs, Hillsboro, OR 97124 USA. This work was done when he was with the School of Electrical and Computer
Engineering, Georgia Institute of Technology, Atlanta, GA 30332 USA (email: lliang@gatech.edu)
}
}
\maketitle

\vspace{-1.7em}

\begin{abstract}
In this article, we develop an end-to-end wireless communication system using deep neural networks (DNNs), in which DNNs are employed to perform several key functions, including encoding, decoding, modulation, and demodulation.
However, an accurate estimation of  instantaneous channel transfer function, \emph{i.e.}, channel state information (CSI), is needed in order for the transmitter DNN to learn to optimize the receiver gain in decoding.
This is very much a challenge since CSI varies with time and location in wireless communications and is hard to obtain when designing transceivers.
We propose to use a conditional generative adversarial net (GAN) to represent channel effects and to bridge the transmitter DNN and the receiver DNN so that the gradient of the transmitter DNN can be back-propagated from the receiver DNN.
In particular, a conditional GAN is employed to model the channel effects in a data-driven way, where the received signal corresponding to the pilot symbols is added as a part of the conditioning information of the GAN.
To address the curse of dimensionality when the transmit symbol sequence is long,  convolutional layers are utilized.
From the simulation results, the proposed method is effective on additive white Gaussian noise (AWGN) channels, Rayleigh fading channels, and frequency-selective channels, which opens a new door for building data-driven DNNs for end-to-end communication systems.

\end{abstract}


\section{Introduction}

\begin{figure}[!t]
\centering
\includegraphics[width=0.8\linewidth]{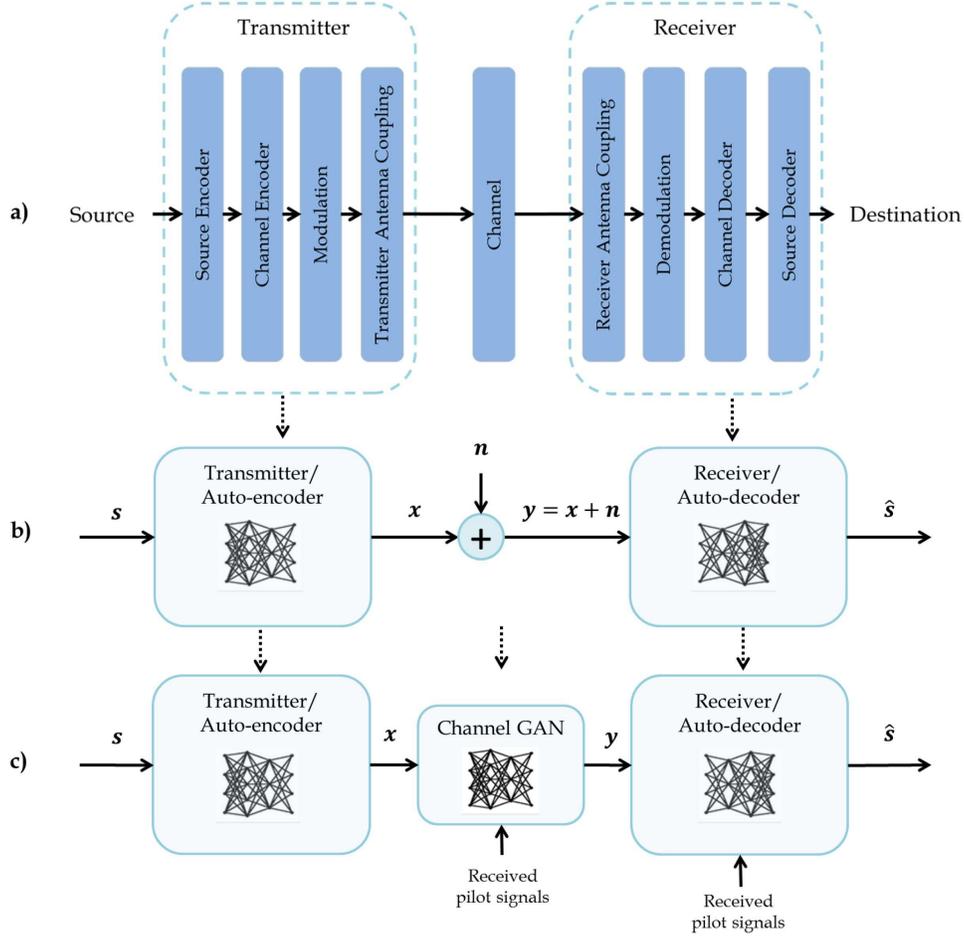}
\caption{Structures of a conventional wireless communication system and end-to-end learning based communication systems: a) conventional wireless communication system; b) end-to-end communication system based on auto-encoder, where the noise is added to the embedding to simulate the channel \cite{physical_layer}; c) proposed end-to-end communication system, where channel GAN is used to model the channel effects.} \label{fig:commblocks}
\end{figure}

In a traditional wireless communication system shown in Fig. \ref{fig:commblocks}.a), the data transmission entails multiple signal processing blocks in the transmitter and the receiver.
While the technologies in this system are quite mature, individual blocks therein are separately designed and optimized, often with different assumptions and objectives, making it difficult, if not impossible, to ascertain global optimality of the system.
In addition, the channel propagation is expressed as an assumed mathematical model embedded in the design.
The assumed model may not correctly or accurately reflect the actual transmission scenario, thereby compromising the system performance.

On the contrary, the learning based data-driven methods provide a new way for handling the imperfection of the assumed channel models \cite{PHY_Qin}.
Recently, deep learning has been applied to refine the traditional block-structure communication systems, including the multiple-input and multiple-output (MIMO) detection \cite{MIMO_Det, HE_MIMO}, channel decoding \cite{Decoding_DNN, Decoding_RNN, Polar, Polar_par, Chuan}, and channel estimation \cite{Channel_DNN, XiaoGuo}.
In addition, deep learning based methods have also shown impressive improvement by jointly optimizing the processing blocks, including joint channel estimation and detection \cite{Hao}, joint channel encoding and source encoding \cite{Source_Channel_Coding}.

Besides enhancing the traditional communication blocks, deep learning provides a new paradigm for future communication systems. As a pure data-driven method, the features and the parameters of a deep learning model can be learned directly from the data, without handcraft or ad-hoc designs, by optimizing an end-to-end loss function. Inspired by this methodology, end-to-end learning based communication systems have been investigated in several prior works \cite{physical_layer,Air, ofdm_end_to_end, RL_E2E, Raj}, where both the transmitter and the receiver are represented by deep neural networks (DNNs) and can be interpreted as an auto-encoder and an auto-decoder, respectively, as shown in Fig ~\ref{fig:commblocks}.b).

From Fig. \ref{fig:commblocks}b), the transmitter learns to encode the transmitted symbols into encoded data, $\mathbf{x}$, which is then sent to the channel， while the receiver learns to recover the transmitted symbols based on the received signal, $\mathbf{y}$, from the channel.
As a result, the traditional communication modules at the transmitter, such as the encoding and modulation, are replaced by a DNN while the modules in the receiver, such as the decoding and the demodulation, are replaced by another DNN.
The weights/parameters of the two DNNs are trained in a supervised learning manner to optimize the end-to-end recovery accuracy.
The robustness to noise can be obtained by adding  noise to the hidden layers to simulate the effect of the wireless channels.

However, several critical challenges in the learning based end-to-end communication system need to be addressed in order to apply this framework to various wireless channels.
As is well known, the weights of the DNN are usually updated using stochastic gradient descent (SGD) with the computed error gradients propagated from the output layer back to the input layer.
When the channel transfer function, $\mathbf{y} = f_\mathbf{h}(\mathbf{x})$, is not available, the back-propagation of the gradients from the receiver DNN to the transmitter DNN is blocked, preventing the overall learning of the end-to-end system.
The channel transfer function may be assumed, but any such assumption would bias the learned weights, repeating the pitfalls caused by the likely discrepancy between the assumed model and the actual channel.
In addition, in real communication systems, an accurate instantaneous CSI is hard to obtain in advance due to
the various inherent uncertainties of wireless channels, such as channel noise and being time-varying.
These uncertainties are often unknown or cannot be expressed analytically.

Another key challenge of the end-to-end paradigm is the curse of dimensionality during the training when the transmitted symbol sequence is long.
The code block size in a communication system needs to be long enough to ensure a sufficient coding gain.
However, as the size of possible codewords grows exponentially with the code block size, the portion of the unseen codewords during training will significantly increase accordingly.
Previous works on the learning based decoding \cite{Polar} show that the decoding performance of the DNN on the unseen codewords is still poor even if nearly  $90\%$ of the codewords have been included in the training of the DNN. Therefore, almost all the previous works on the end-to-end paradigm are concentrating on examples with a small block size, such as the Hamming codes (7,4) \cite{physical_layer, Channel_GAN}.
As a result, it is desirable to develop a channel agnostic end-to-end communication system based on deep learning, where different types of channel effects can be automatically learned without knowing the specific channel transfer function and the block-length remains long enough to be practical.



In this article, we develop a channel agnostic end-to-end communication system to address the challenges, where the distributions of channel output are learned through a conditional generative adversarial net (GAN) \cite{conditional}, as shown in Fig~\ref{fig:commblocks}.c).
The conditioning information for the GAN to generate samples is the encoded signals from the transmitter along with the received pilot information used for estimating the channel.  By iteratively training the conditional GAN, the transmitter, and the receiver, the end-to-end loss can be optimized in a supervised way.

This channel agnostic end-to-end system provides a new way to optimize communication systems and is applicable to a wide range of wireless channels.
In addition, the convolutional neural network (CNN) is used to overcome the curse of dimensionality and the block-length can be extended from several bits to a couple of hundred bits.
Our main contributions in this article are fourfold.
\begin{itemize}
    \item We are the first to exploit the conditional GAN to model the channel conditional distribution, $p(\mathbf{y}|\mathbf{x})$, so that the channel effects can be learned based on the data instead of expert knowledge.
    \item By adding the pilot symbol as a part of the conditioning information for the time-varying channels, the conditional GAN can generate more specific samples for the current channel.
    \item Based on the learned channel conditional distribution, an end-to-end learning based communication system is developed, where the gradients of the end-to-end loss can be propagated to the transmitter DNN through the conditional GAN.
    \item CNN is employed  for alleviating curse of dimensionality. From the experimental results, the transmitter DNN with convolutional layers can learn to encode the information bits into a high dimensional embedding vector and the code can be efficiently decoded by the receiver DNN.
\end{itemize}

Part of the work has been published in \cite{Channel_GAN}.
Compared with the previous work, we have made two significant improvements.
First, we introduce a convolutional layers so that this approach can be extended from several bits to a couple of hundred bits. Second, we apply our framework to more practical wireless channels, such as frequency-selective channels
where there exists inter-symbol interference (ISI).

The rest of the paper is organized as follows. The related works are discussed in Section \ref{sec:Related_Work}.
In Section \ref{sec:GAN}, the conditional GAN based channel modeling approach is introduced. In Section \ref{sec:End2End}, the training for the end-to-end system is presented in detail. In Section \ref{sec:Exp}, the simulation results are presented and the conclusions are drawn in Section \ref{sec:Conclusion}.

\section{Related Works} \label{sec:Related_Work}
Our proposed method is closely related to GANs, end-to-end learning based communication systems, and learning based decoders.
In this section, previous works in the related topics are briefly reviewed.

\subsection{GANs and Conditional GANs}

GAN has been proposed in \cite{GAN} as a generative framework, where a generator and a discriminator are competing with each other in the training stage.
By the feedback of the discriminator, the generator improves its ability to generate samples that are similar to the real samples.
GAN is most widely used in computer vision. Much of the recent GAN research is focusing on improving the quality of the generated images \cite{BIGGAN}.

In order to generate samples with a specific property, a conditional GAN is proposed based on the GAN framework, where the context information is added to the generator and the discriminator.
Originally, the condition added is the label information so that the generator can generate sample data given a particular category.
Nowadays, conditional GAN is widely used in changing the style and the content of the input \cite{SRGAN,Cycle_GAN}.
For instance, GAN has been utilized to generate high-resolute images from low-resolution images \cite{SRGAN}.


Apart from application in computer vision, recently GAN has been exploited to model the channel effects of additive white Gaussian noise (AWGN) channels \cite{void}, similar to our work.
However, our approach can be applied to more realistic time-varying channels by using conditional GAN, which employs the received pilot information as a part of the condition information when generating the channel outputs.

\subsection{DNN based End-to-End Communications}

An end-to-end learning system has been proposed in \cite{physical_layer} and has been shown to have a similar performance as the traditional approaches with block structures under the AWGN condition.
In \cite{Air}, the end-to-end method has been extended to handle various hardware imperfection.
In \cite{ofdm_end_to_end}, an end-to-end learning method is adopted within the orthogonal frequency-division multiplexing (OFDM) system.
In \cite{CNN_Modulation},  CNN is employed for modulation and demodulation, where improved results have been shown for very high order modulation.
In addition, source coding can also be considered as a part of the end-to-end communication system for transmitting text or image \cite{Source_Channel_Coding}.

Training the end-to-end communication system without channel models has been investigated recently.
A reinforcement learning based framework has been employed in \cite{RL_E2E} to optimize the end-to-end communication system without requiring the channel transfer function or CSI, where the channel and the receiver are considered as the environment when training the transmitter.
The recovery performance at the receiver is considered as the reward, which guides the training of the transmitter.
In \cite{Raj}, a model-free end-to-end learning method has been developed based on stochastic perturbation methods.
However, both works focus on small block length. For example, blocks of eight information bits are used in \cite{RL_E2E}.
How to extend a large block size and how to model the unknown channel using a data-driven approach are still open problems, which are addressed in our proposed approach.

\subsection{Learning based Decoders}
Our proposed method is also closely related to learning based encoding and decoding.
Learning based approach has been utilized in improving decoding performance for a long time.  It dates back to 1990s when several attempts had been made to decode the codes with the recurrent neural network (RNN) \cite{Decoding_RNN_1999}.

With the widely used deep learning approaches, DNN has been utilized in decoding with a wide range of applications.
In \cite{Polar}, a fully connected neural network is trained for decoding short Polar codes. The performance of decoding is similar to maximum likelihood decoding. But they find it difficult to break through the curse of dimensionality.
In order to train long codewords, a partition method has been employed in \cite{Polar}.
Moreover, there have been several trials to incorporate some prior information in the decoding process.
RNN is used in \cite{Decoding_RNN} for decoding the convolutional and turbo codes.
In \cite{Decoding_DNN}, the traditional belief-propagation decoding algorithm is extended as deep learning layers to decode linear codes.

One of the interesting findings from our experiment is that the curse of dimensionality can be mitigated when the CNN is used to learn the encoding and decoding modules simultaneously rather than just learning to decode  human-designed codewords.

\section{Modeling Channel with Conditional GAN} \label{sec:GAN}

An end-to-end communication system learns to optimize DNNs for the transmitter and the receiver.
However, the back-propagation, which is used to train the weights of DNNs, is blocked by the unknown CSI, preventing the overall learning of the end-to-end system.
To address the issue, we use a conditional GAN to learn the channel effects and to act as a bridge for the gradients to pass through.
By the conditional GAN, the output distribution of the channel can be learned in a data-driven manner and therefore many complicated effects of the channel can be addressed.
In this section, we introduce the conditional GAN and discuss how to use it to model the channel effects.

\subsection{Conditional GAN}


GAN \cite{GAN} is a new class of generative methods for distribution learning, where the objective is to learn a model that can produce samples close to some target distribution, $p_{data}$. In our system, a GAN is applied to model the distribution of the channel output and the learned model is then used as a surrogate of the real channel when training the transmitter so that the gradients can pass through to the transmitter.

\begin{figure}[!t]
\centering
\includegraphics[width=0.5\linewidth]{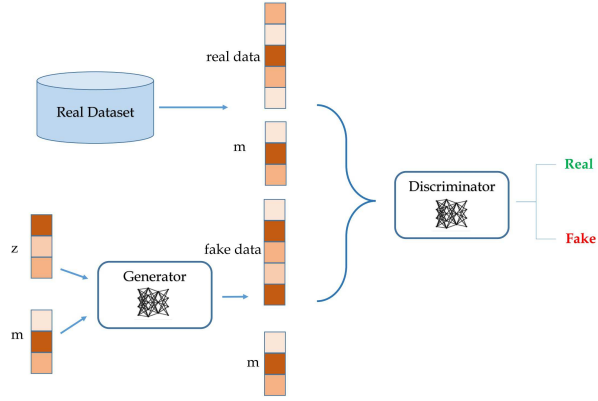}
\caption{Structure of conditional GAN.} \label{fig:CGAN}
\end{figure}

As shown in Fig.~\ref{fig:CGAN}, a min-max two players game is introduced in GAN between a generator, $G$, and a discriminator, $D$, both represented by DNNs. The discriminator, $D$, learns to distinguish between the data generated by the generator and the data from the real dataset while the generator, $G$, learns to generate samples to fool the discriminator into making mistakes.

During the training, the generator maps an input noise, $\mathbf{z}$, with prior distributions, $p_z(\mathbf{z})$, to a sample. Then the samples from the real data and those generated from $G$ are collected to train $D$, to maximize the ability to distinguish between the two categories. If the discriminator can successfully classify the samples of the two sources, then its success will be used to generate feedback to $G$, so that the generator can learn to produce samples more similar to the real samples. The training procedure will end upon reaching an equilibrium, where the discriminator, $D$, can do no better than random guessing to distinguish the real samples and the generated fake samples.

Denote the parameter sets of the generator, $G$, and the discriminator, $D$, as $\mathcal{G}$ and $\mathcal{D}$, respectively, and the objective for optimization is

\begin{equation}
\begin{split}
\min_{\mathcal{G}}\max_{\mathcal{D}} V(D, G) =& E_{\mathbf{x}\sim p_{data}(\mathbf{x})} [\log(D_{\mathcal{D}}(\mathbf{x}))] \\
&+ E_{\mathbf{z}\sim p_z(\mathbf{z})}[\log(1 - D_{\mathcal{D}}(G_{\mathcal{G}}(\mathbf{z})].
\end{split}
\end{equation}

The objective of the discriminator, $D$, is to give a high value when the input belongs to the real dataset and a low one when the input is generated by the generator, $G$, while the objective of generator, $G$, is to maximize the output of the discriminator, $D$, given the generated samples, $G(\mathbf{z})$.

The GAN can be extended to a conditional model if both the generator, $G$, and the discriminator, $D$, are conditioned on some extra information, $\mathbf{m}$, as in Fig. \ref{fig:CGAN}. We only need to feed the conditioning information, $\mathbf{m}$, into both the generator and discriminator as the additional input. Therefore, the output of the $G$ will be $G(\mathbf{x}|\mathbf{m})$ and the output of $D$ will be $D(\mathbf{x}|\mathbf{m})$.  The min-max optimization objective becomes
\begin{equation}
\begin{split}
\min_{\mathcal{G}}\max_{\mathcal{D}} V(D, G) &= E_{\mathbf{x}\sim p_{data}(\mathbf{x})} [\log(D_{\mathcal{D}}(\mathbf{x}|\mathbf{m}))] \\
&+ E_{\mathbf{z}\sim p_z(\mathbf{z})}[\log(1 - D_{\mathcal{D}}(G_{\mathcal{G}}(\mathbf{z}|\mathbf{m})].
\end{split}
\end{equation}
The conditional GAN is employed in our end-to-end communication system to model the channel output distribution with the given conditioning information on the encoded signal and the received pilot data.

 \begin{figure}[!t]
  \centering
\includegraphics[width=0.5\linewidth]{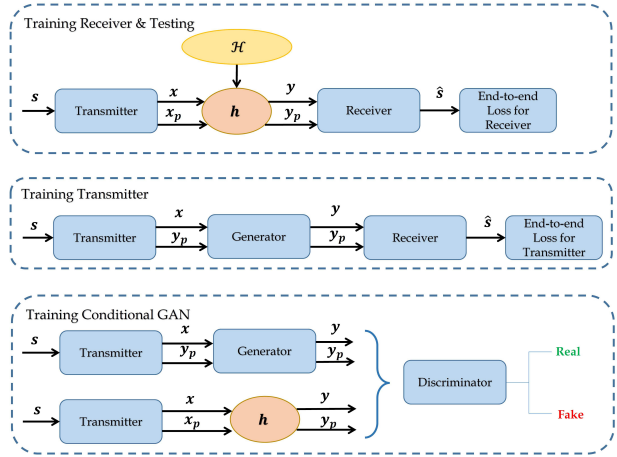}
\caption{Training and testing of the end-to-end system.} \label{fig:E2E}
\end{figure}

\subsection{Modeling Channels}

Since the channel output, $\mathbf{y}$, for given input, $\mathbf{x}$, is determined by the conditional distribution, $p(\mathbf{y}|\mathbf{x})$, a conditional GAN can be employed for learning the output distribution of a channel by taking $\mathbf{x}$ as the condition information.
The generator will try to produce the samples similar to the output of the real channel while the discriminator will try to distinguish data coming from the real channel and the data coming from the generator.

The instantaneous CSI, $\mathbf{h}$, is regarded as a sample from a large channel set $\mathcal{H}$ and is also vital for coherent detection of the transmit symbols at the receiver. In order to obtain the CSI, a common practice is to send some pilot information to the receiver so that the channel information is inferred based on the received signal corresponding to the pilot symbols, $\mathbf{y_p}$.  In our proposed method, the received signal corresponding to the pilot symbols, $\mathbf{y_p}$, is added as a part of the conditioning information so that the output samples follow the distribution of $\mathbf{y}$ given the input $\mathbf{x}$ and the received pilot data, $\mathbf{y_p}$.



\subsection{Convolutional Layers based Channel GAN}

The convolutional layer has been introduced to efficiently extract features for images based on their shared-weight architecture and translational invariance characteristics \cite{Le_CNN}.
In a fully connected layer, each neuron is connected to all neurons in the previous layer.
In contrast, in a convolutional layer, each neuron is only connected to a few nearby neurons in the previous layer, which is called the receptive field of this neuron, and the same set of weights is shared for all neurons in a layer.
Inspired by the convolutional codes, where the encoding process can be represented by a convolutional transform, we use hierarchical one-dimensional convolutional layers in the channel GAN as well as the DNNs used in the transmitter and the receiver.

Denote ${u}^{(i)}{[n]}$ as the output of the $n$-th neuron in the $i$-th layer of a DNN. For a fully connected layer, the output of the $n$-th neuron in the $i$-th layer is
\begin{equation*}
    {u}^{(i)}{[n]} = \sigma(\sum_k w_{nk}^{(i)}{u}^{i-1}[k]),
\end{equation*}
where $\sigma(\cdot)$ is an activation function and $w_{nk}^{(i)}$ is the weight connected the $k$-th neuron in the $(i-1)$-th layer and the $n$-th neuron in the $i$-th layer. $w_{nk}^{(i)}$ is different for different $i$, $n$, or $k$.
Therefore, if there are $N_i$ neurons in the $i$-th layer, there will be $N_iN_{i-1}$ weights in total to fullly connect the $(i-1)$-th layer to the $i$-th layer.

On the other hand, for a convolutional layer, the output of the $n$-th neuron in the $i$-th layer will be
\begin{equation*}
    {u}^{(i)}{[n]} = \sigma(\sum_{k=1}^Lw_k^{(i)}{u}^{(i-1)}[n-k]),
\end{equation*}
where $w_k^{(i)}$ is the coefficient for the convolution and is same for different $n$'s in the $i$-th layer. There are $L$ weights in total ($L$ is usually much smaller then $N_i$ or $N_{i-1}$) to implement $N_iL$ connections between the $(i-1)$-th and the $i$-th layers. In brief, compared with a fully connected DNN, the convolutional nerual network has fewer connections between adjacent layes and much fewer weights to train, which will reduce the complexity and significantly improve the convergence speed of training.

Apart from being easier to train, CNN has two additioinal merits in the end-to-end communication system.
First, the curse of dimensionality can be alleviated by the usage of convolutional layers.
When both the transmitter and the receiver are represented by CNNs, the codes learned by a CNN are more easily  recovered at the receiver than the conventional hand designed codes.
Second, it is appropriate to employ convolutional layers to deal with the ISI channels
since the effect of the channel can be expressed by the convolutional operation in the ISI channel.

\section{End-to-End Communication System} \label{sec:End2End}
As stated in the introduction, the end-to-end communication paradigm can be interpreted as a deep auto-encoder framework.
With the conditional GAN, the gradients can be back-propagated to the transmitter even if channels are known. In this section, the proposed framework is first introduced and the training procedures for each module are presented in detail.

\subsection{System Overview}
As in Fig.~\ref{fig:commblocks}.b) the auto-encoder learns to map $N$ information bits, $ \mathbf{s} \in \{0, 1\}^N $ into a fixed length embedding of length $K$, $\mathbf{x} \in \mathbb{R}^K$, and sends the embedding to the channel while the auto-decoder learns to recover the original information according to the received signal $\mathbf{y}$ from the channel. The distance between the original information bits, $\mathbf{s}$, and the recovered information, $\hat{\mathbf{s}} \in [0, 1]^N$, will be calculated. Here, the binary cross-entropy loss is used to measure the distance, which can be expressed as
\begin{equation}
L = \sum_{n=1}^{M} −(s_n \log(\hat{s}_n)+(1 - s_n) \log(1 - \hat{s}_n)). \label{equ:loss}
\end{equation}
where $s_n$ and $\hat{s}_n$ represent the $n$th elements of $\mathbf{s}$ and $\mathbf{\hat{s}}$, respectively.

The training and testing of the proposed end-to-end communication system are shown in Fig.~\ref{fig:E2E}.
To obtain training data set, the information bits, $\mathbf{s}$, are randomly generated and the instantaneous CSI is sampled randomly from the channel set.
Due to different objectives in modules, the transmitter, the receiver, and the channel generator in the conditional GAN can be trained iteratively based on the training data. When training one component, the parameters of the others remain fixed.
The object is to minimize the end-to-end loss when training the receiver and the transmitter.
It is to minimize the min-max optimization objective when training the conditional GAN for generating the channel. In the testing stage, the end-to-end reconstruction performance is evaluated on the learned transmitter and receiver with real channels.

\subsection{Training Receiver}
At the receiver, a DNN model is trained for recovering the transmitted signal $\mathbf{s}$, where the input is the received signals corresponding to the transmitted data, $\mathbf{y}$, while the output is the estimation $\mathbf{\hat{s}}$.
By comparing the  $\mathbf{s}$ and $\mathbf{\hat{s}}$, the loss function can be calculated based on (3). The receiver can be trained easily since the loss function is computed at the receiver and thus the gradients of the loss can be easily obtained.
For the time-varying channels, by directly put the received signal, $\mathbf{y}$, and the receive pilot data, $\mathbf{y_p}$, together as the input, the receiver can automatically infer the channel condition and perform the channel estimation and detection simultaneously without explicitly estimating the channel, as we have discussed in \cite{Hao}.

\subsection{Training Transmitter}

With the channel generator as a surrogate channel, the training of the transmitter will be similar to that of the receiver.
During the training, the transmitter, the generator, and the receiver can be viewed as a whole DNN.
The output of the transmitter is the values of the last hidden layer in the transmitter.
The end-to-end cross-entropy loss is computed at the receiver as in (3), and the gradients are propagated back to the transmitter through the conditional GAN.
The weights of the transmitter will be updated based on SGD while the weights of the conditional GAN and the receiver remain fixed.
The transmitter can learn the constellation of the embedding, $\mathbf{x}$, so that the received signal can be easily detected at the receiver.

\subsection{Training Channel GAN}
The conditional GAN is trained by iteratively training the generator and discriminator so that the min-max point can be found.
The parameters of one model will be fixed while training the other.
The channel generator is trained with the discriminator together. With the learned transmitter, the real data can be obtained with the encoded signal from the transmitter going through the real channel while the fake data is obtained from the encoded data going through the channel generator. The objective function for optimization is as shown in (2).

\section{Experiments} \label{sec:Exp}

In this section, the implementation details of the end-to-end learning based approach are provided and the simulation results are presented.
For several types of most commonly used channels, the channel GAN has shown the ability to model the channel effects in a data-driven way.
In addition, the end-to-end communication system, which is built on the channel GAN, can achieve similar or better results even when the channel information is unknown when training and optimizing the transmitter and the receiver.

\subsection{Experimental Settings}

\subsubsection{Implementation Details}
Two types of DNN models are designed in our experiments. One is fully connected networks (FCN) and the other is the CNN.
The FCN is used for a small block size and the CNN is used in the a large block size to avoid the curse of dimensionality.
The parameters of the FCN and CNN are shown in Table \ref{table:FCN} and Table \ref{table:CNN}, respectively.
The weights of both models are updated by Adam \cite{ADMM} and the batch size for training is 320.

\begin{table}
\centering
\caption{Model Parameters of FCN} \label{table:FCN}
\begin{tabular}{|c|c|}
\hline
\rowcolor{gray}
Parameters & Values \\ \hline
Transmitter hidden layers & {32, 32} \\ \hline
Learning rate & 0.001 \\ \hline
Receiver hidden layers & {32, 32} \\ \hline
Learning rate & 0.001 \\ \hline
Generator hidden layers & {128, 128, 128} \\ \hline
Discriminator hidden layers & {32, 32, 32} \\ \hline
Learning rate & 0.0001 \\ \hline
\end{tabular}\label{tab:Model}
\end{table}

\begin{table}
\centering
\caption{Model Parameters of CNN} \label{table:CNN}
\begin{tabular}{|c|c|c|}
\hline
\rowcolor{gray}
Type of layer & Kernel size/Annotation & Output size  \\ \hline
\multicolumn{3}{|c|}{\textbf{Transmitter}} \\\hline
Input  & Input layer & $K \times 1$  \\ \hline
Conv+Relu & 5 & $K \times 256$  \\ \hline
Conv+Relu & 3 &  $K \times 128 $ \\ \hline
Conv+Relu & 3 & $K \times 64 $ \\ \hline
Conv & 3 & $K \times 2 $ \\ \hline
Normalization & Power normalization & $K \times 2 $ \\ \hline
\multicolumn{3}{|c|}{\textbf{Receiver}} \\\hline
Conv+Relu & 5 &  $K \times 256$\\ \hline
Conv+Relu & 5 &  $K \times 128$ \\ \hline
Conv+Relu & 5 &  $K \times 128$ \\ \hline
Conv+Relu & 5 &  $K \times 128$ \\ \hline
Conv+Relu & 5 &  $K \times 64$ \\ \hline
Conv+Relu & 5 &  $K \times 64$ \\ \hline
Conv+Relu & 5 &  $K \times 64$ \\ \hline
Conv+Sigmoid & 3 & $K \times 1 $\\ \hline
\multicolumn{3}{|c|}{\textbf{Generator}} \\\hline
Conv+Relu & 5 & $K \times 256$  \\ \hline
Conv+Relu & 3 &  $K \times 128 $ \\ \hline
Conv+Relu & 3 & $K \times 64 $ \\ \hline
Conv & 3 & $K \times 2 $ \\ \hline
\multicolumn{3}{|c|}{\textbf{Discrimator}} \\\hline
Conv+Relu & 5 & $K \times 256$  \\ \hline
Conv+Relu & 3 &  $K \times 128 $ \\ \hline
Conv+Relu & 3 & $K \times 64 $ \\ \hline
Conv+Relu & 3 & $K \times 16 $ \\ \hline
FC +Relu  & 100 &  100 \\ \hline
FC+Sigmoid & 1 &  1  \\ \hline
\end{tabular}\label{tab:Model}
\end{table}

\subsubsection{Channel Types}
Three types of channels are considered in our experiments, \emph{i.e.}, AWGN channels, Rayleigh channels, and frequency-selective multipath channels.
In an AWGN channel, the output of the channel, $\mathbf{y}$, is the summation of the input signal, $\mathbf{x}$, and Gaussian noise, $\mathbf{w}$, that is, $\mathbf{y} = \mathbf{x} + \mathbf{w}$.
Rayleigh fading is a reasonable model for narrow-band wireless channels when many objects in the environment scatter the radio signal before arriving at the receiver.
In a Rayleigh channel, the channel output is determined by $y_n = h_n \cdot x_n + w_n$, where $h_n \sim \mathcal{CN}(0,1)$.
The channel coefficient $h_n$ is time-varying and is unknown when design transceivers.
Therefore, channel estimation is required to get the instantaneous CSI, for receiver to detect the transmitter information.

With frequency-selective channels, radio signal propagate via multiple paths, which differ in amplitudes, phases, and delay times, and cause undesired frequency-selective fading
and time dispersion of the received signal.
The baseband complex channel impulse response can be expressed as

\begin{equation*}
    h(t) = \sum_{k=0}^{K_p} {b_k e^{j\theta_k} p(t - \tau_k)},
\end{equation*}
where there are $K_p$ pathes in all, $b_k$, $\theta_k$, and $\tau_k$ represent the path gain, the phase shift, and the time delay of the $k$th path, respectively, and $p(t)$ is the shaping pulse in the communication system.
In our simulation, a three-tap channel with equal average power is considered, that is, $\mathbb{E}|b_k|^2 = 1$, and $\tau_k = 0, T, 2T$ with $T$ as the symbol duration.

\subsubsection{Baselines}
The end-to-end learning based communication system is compared with the conventional communication system, which is composed of multiple signal processing modules and each module is designed based on the prior knowledge on the channel. The bit-error rate (BER) and block-error rate (BLER) are compared under each type of channel.
In our baseline system, 4 QAM is used as the modulation and the Hamming code or convolutional codes are used.
For the convolutional codes, the Viterbi algorithm \cite{viterbi} is used for the maximum a posteriori probability (MAP) decoding.
A commonly used example of a convolutional code, rate-1/2 recursive systematic convolutional (RSC) code, is adopted.
OFDM is utilized to deal with the ISI in the frequency-selective multi-path channel.

\subsection{Modeling the Channel Effects}



\begin{figure}[ht]
  \begin{subfigure}[b]{0.5\linewidth}
    \centering
    \includegraphics[width=1\linewidth]{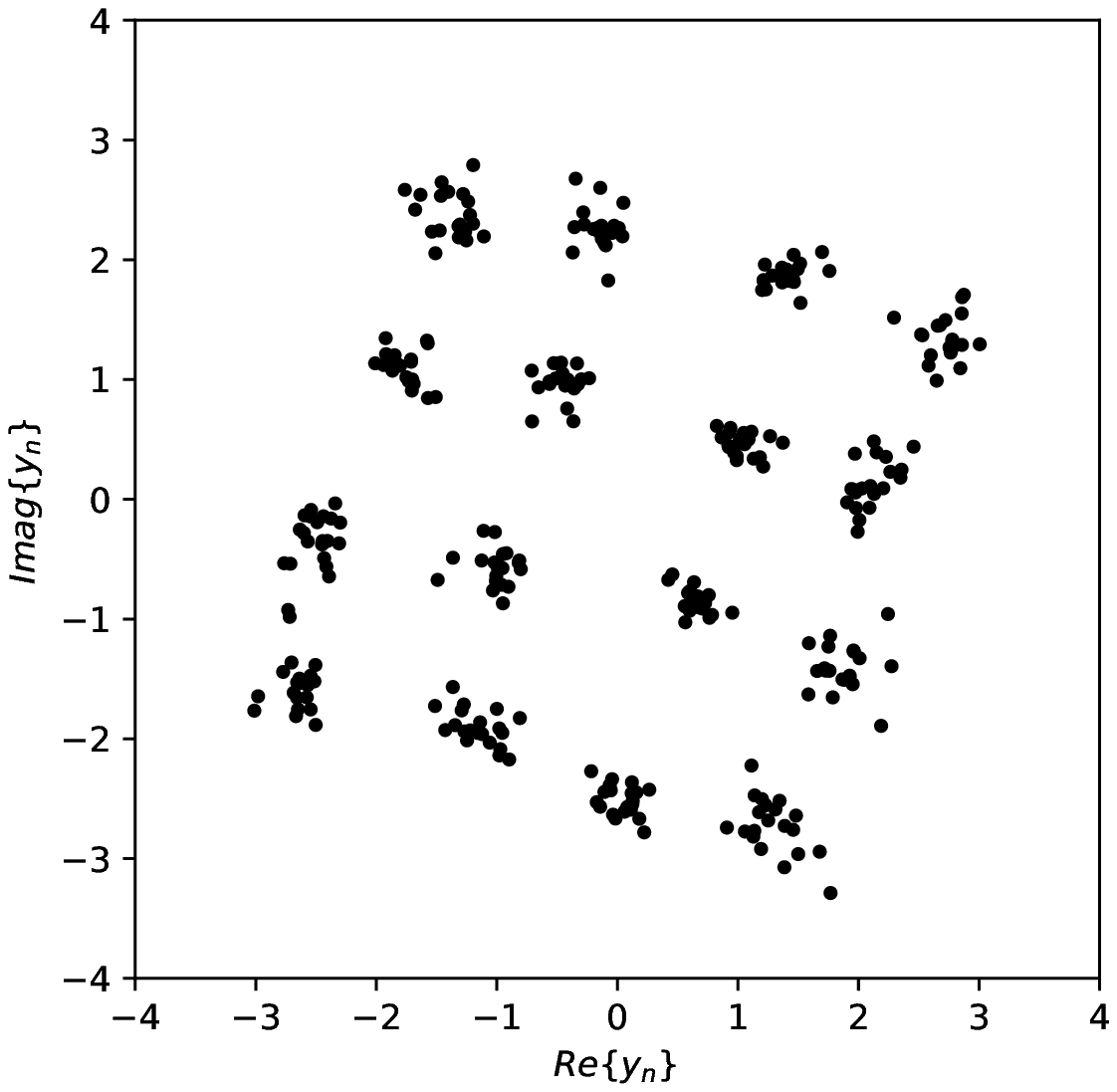}
  \end{subfigure}
  \begin{subfigure}[b]{0.5\linewidth}
    \centering
     \includegraphics[width=1\linewidth]{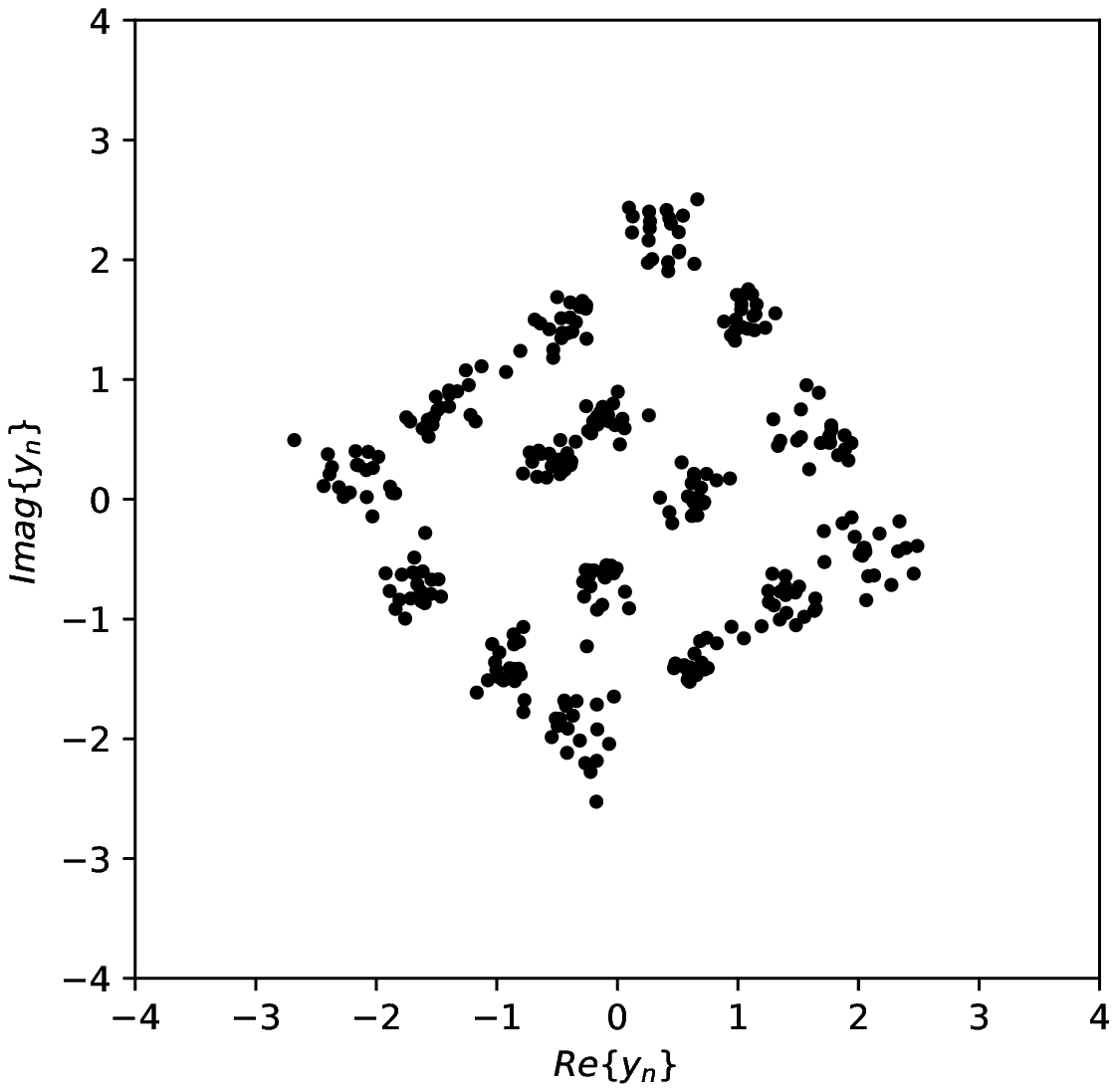}
   \end{subfigure}
   \caption{Signal constellations at the output of a Rayleigh channel represented by a conditional GAN.}
   \label{fig:Ray_mod}
 \end{figure}

We use FCN to model the effects of Rayleigh fading channels. Since Rayleigh fading channels are time-varying, additional conditional information is added to the channel generator and the receiver.
Besides the encoded signal, the received pilot data, $\mathbf{y_p}$, is used as the additional conditional information.
We test the effectiveness of the conditional GAN in learning the distribution of the channel with standard 16 QAM as the encoded symbols.
Fig. \ref{fig:Ray_mod} shows generated samples of a Rayleigh fading channel with different values added to the conditioning information.
From the figure, the conditional GAN is able to produce the samples with different channel gains and phase rotations according to conditioning information.

\subsection{End-to-end Communication System}

Based on the channel GAN, a channel agnostic end-to-end communication system is built on three types of channels, i.e., the AWGN channel, the Rayleigh fading channel, and the frequency-selective multi-path channel.
We compare our channel agnostic end-to-end learning based approach with the traditional methods, which are designed based on the channel transfer functions.

\subsubsection{AWGN Channel}

\begin{figure}[ht]
  \begin{subfigure}[b]{0.5\linewidth}
    \centering
    \includegraphics[width=1\linewidth]{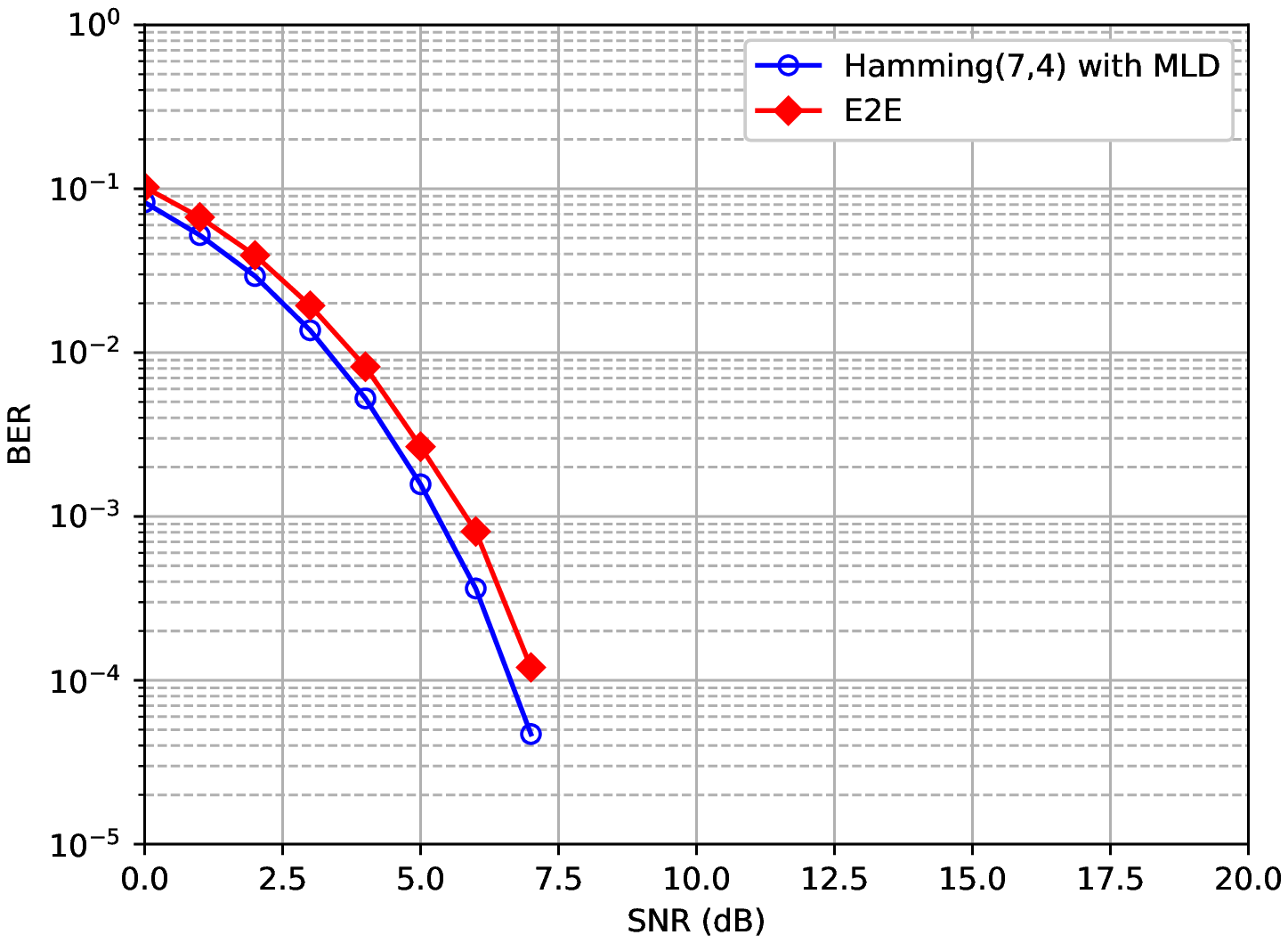}
  \end{subfigure}
  \begin{subfigure}[b]{0.5\linewidth}
    \centering
     \includegraphics[width=1\linewidth]{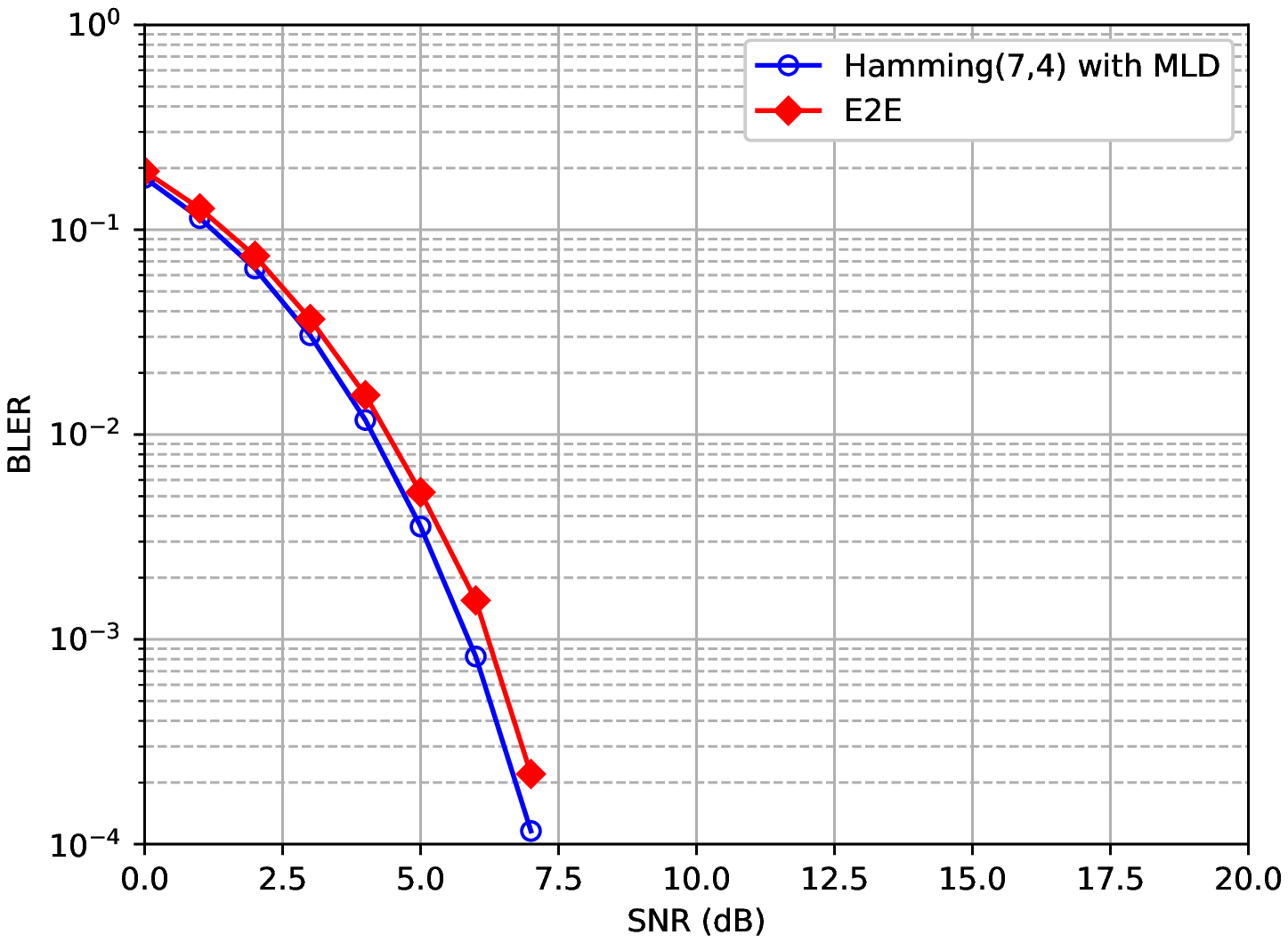}
   \end{subfigure}
   \caption{BER and BLER of a small block size under a AWGN channel.}
   \label{fig:AWGN_ber}
 \end{figure}

We first use FCN for a small block size. The end-to-end recovering performance on the AWGN channel is shown in Fig. \ref{fig:AWGN_ber}. At each time, four information bits are transmitted and the length of the transmitter output is set to be seven. From the figure, the BER and BLER of learning based approach is similar to Hamming (7,4) code with maximum-likelihood decoding (MLD).

\begin{figure}[ht]
  \begin{subfigure}[b]{0.5\linewidth}
    \centering
    \includegraphics[width=1\linewidth]{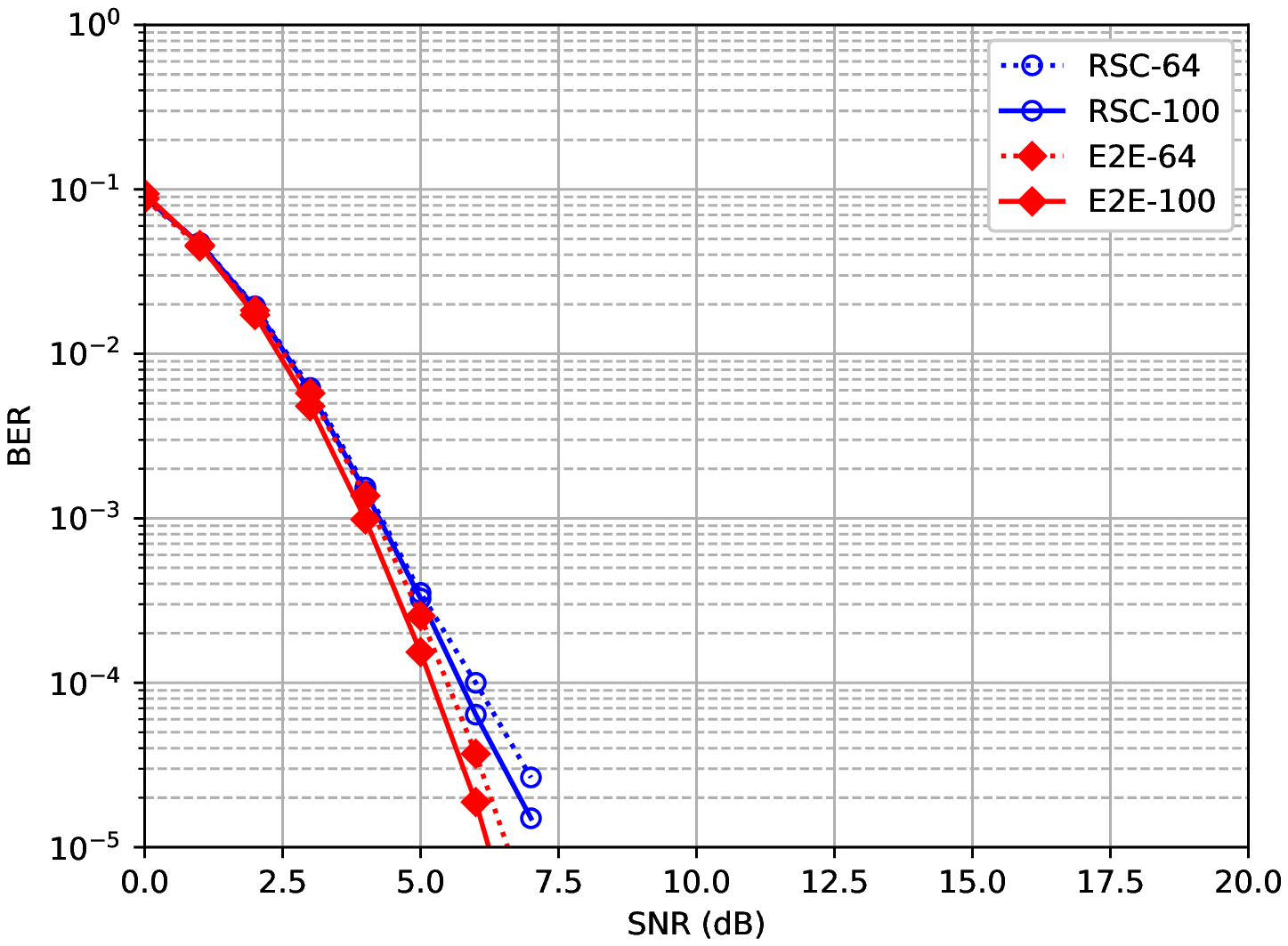}
  \end{subfigure}
  \begin{subfigure}[b]{0.5\linewidth}
    \centering
     \includegraphics[width=1\linewidth]{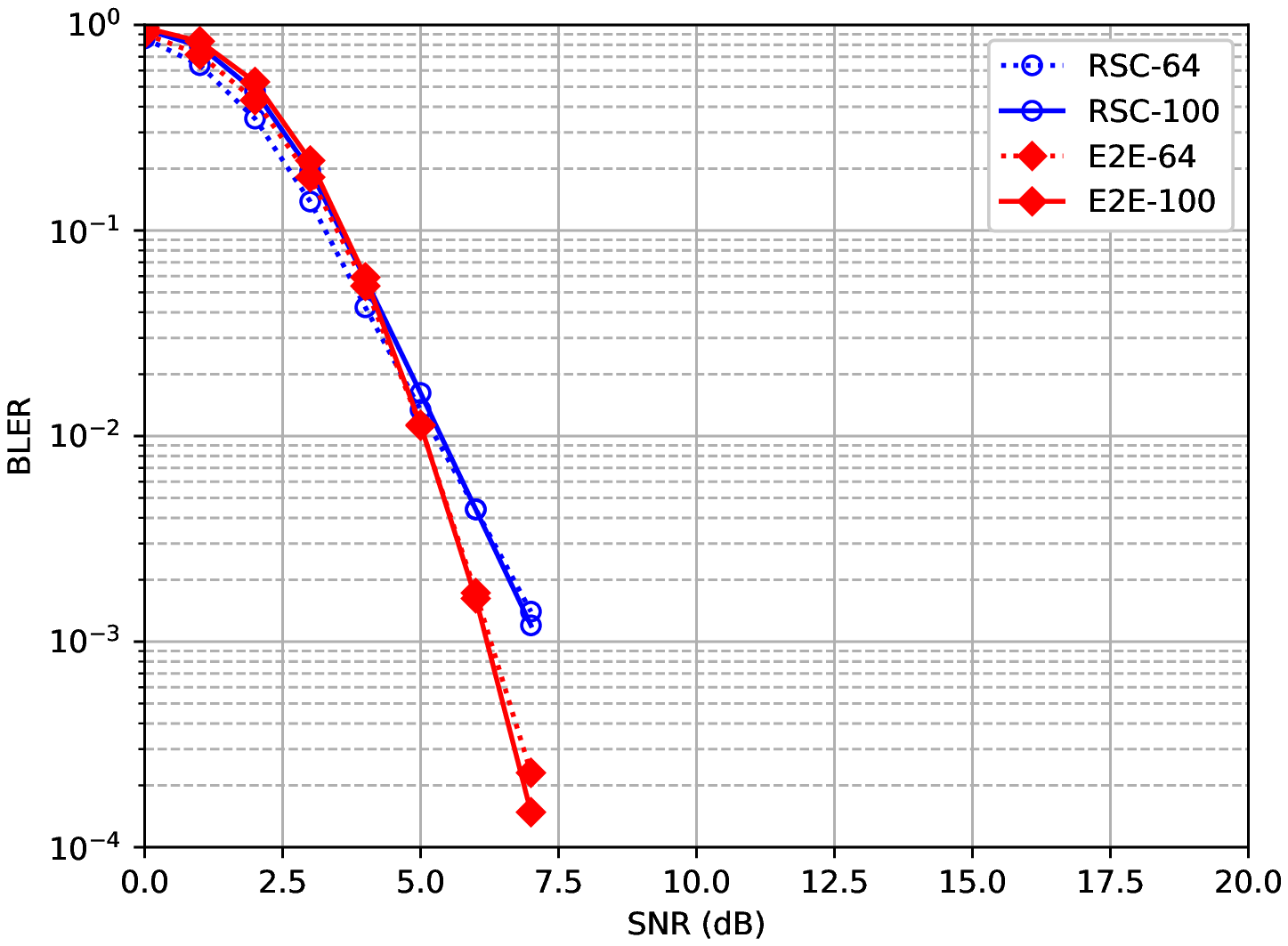}
   \end{subfigure}
   \caption{BER and BLER of a large block size under an AWGN channel.}
   \label{fig:AWGN64}
 \end{figure}

In order to train models with a large block size, CNN is then used to mitigate  the curse of dimensionality.
We first train the CNN under the AWGN channel where the noise is added to the hidden layer directly, as used in \cite{physical_layer}.
The network is trained at  $3$ dB fixed signal-to-noise ratio (SNR) and tested by different SNRs.
Fig. \ref{fig:AWGN64} shows the the BER and BLER curves of proposed end-to-end method with the length of transmit information sequence 64 bits and 100 bits, respectively, which are denoted by ``E2E-64'' and ``E2E-100'', respectively.
From the figure, the performance of the proposed method is similar to RSC in the low SNR area and significantly outperforms RSC in the high SNR area.

\subsubsection{Rayleigh Fading Channel}

\begin{figure}[ht]
  \begin{subfigure}[b]{0.5\linewidth}
    \centering
    \includegraphics[width=1\linewidth]{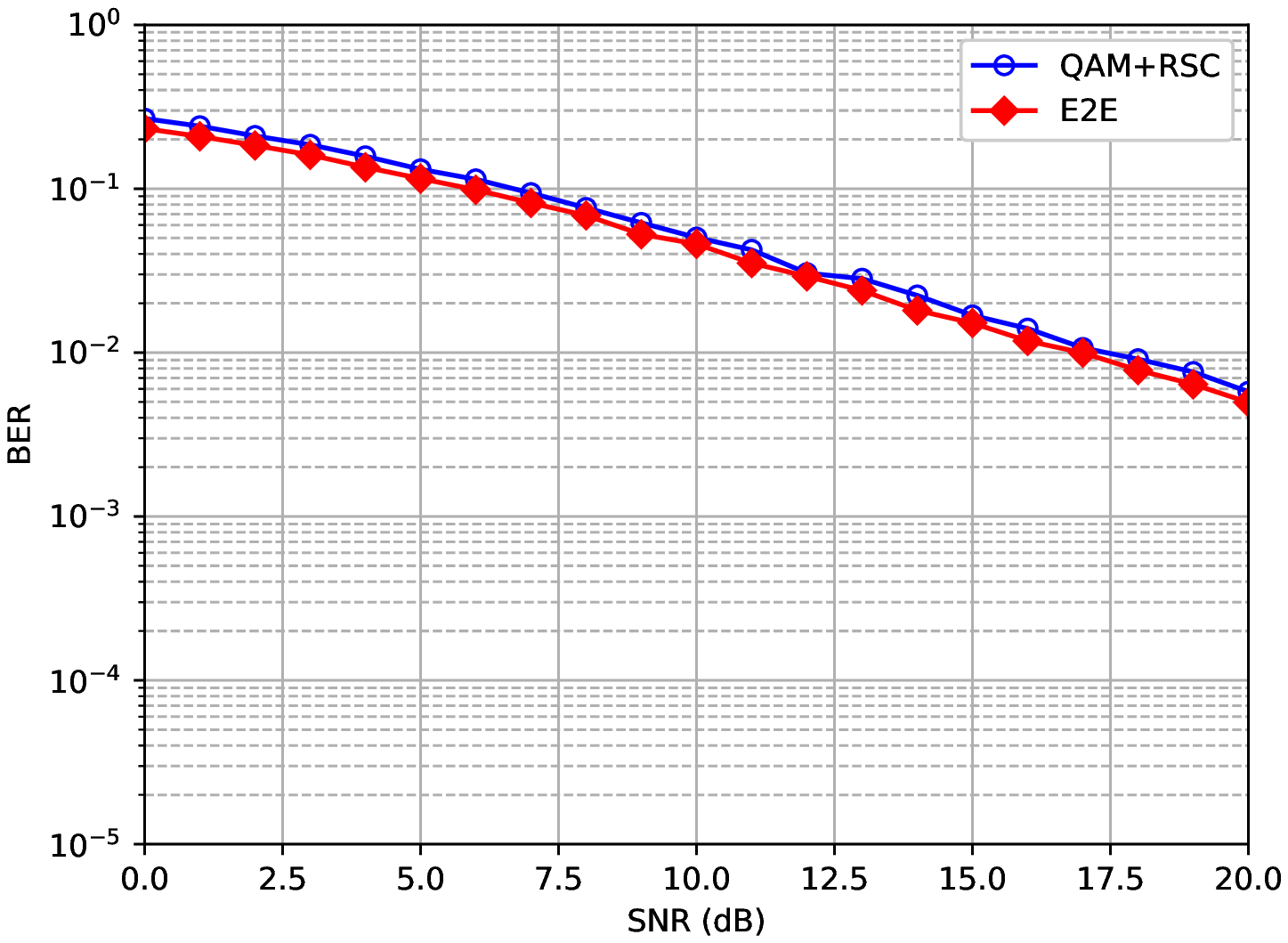}
  \end{subfigure}
  \begin{subfigure}[b]{0.5\linewidth}
    \centering
     \includegraphics[width=1\linewidth]{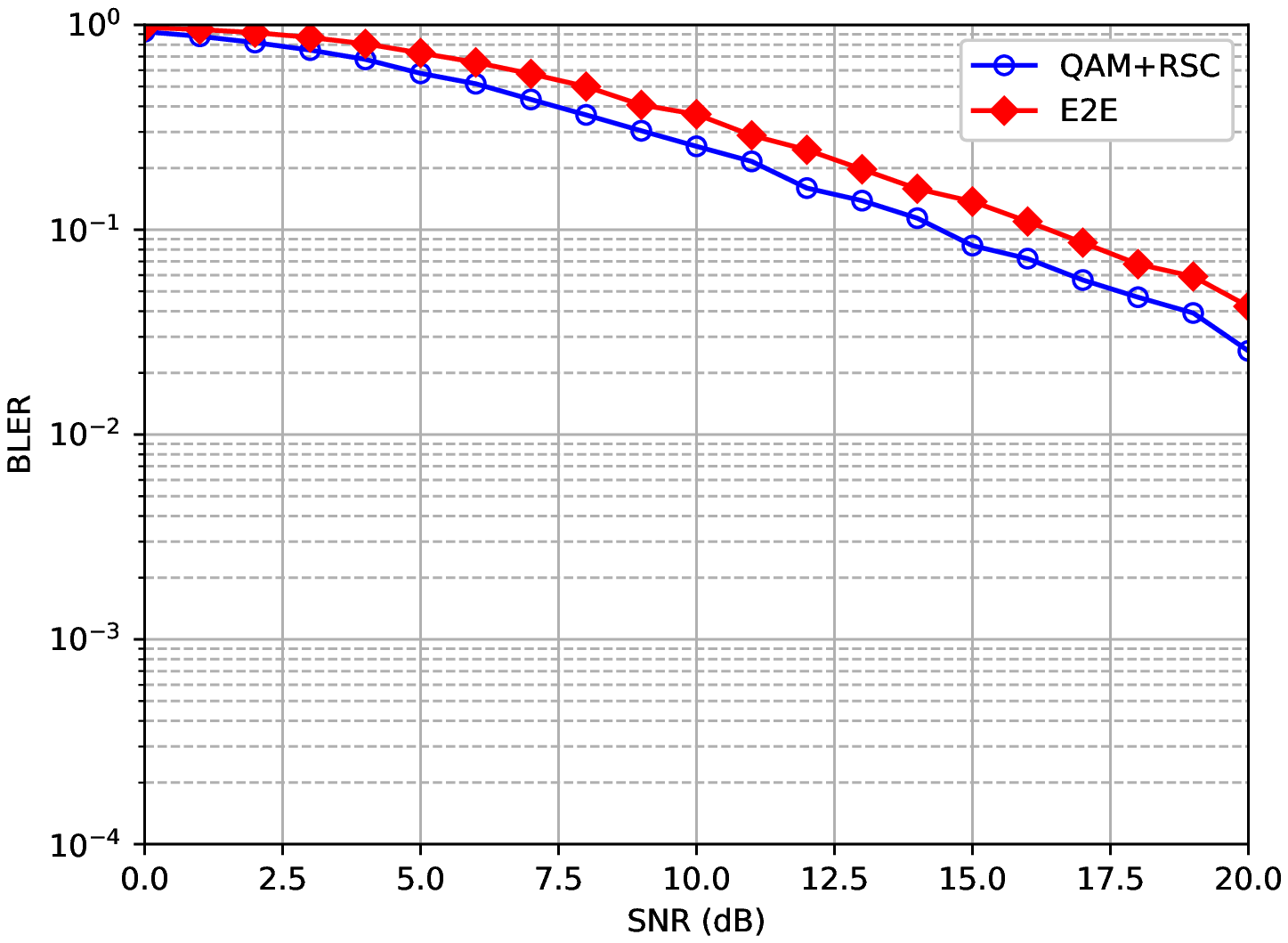}
   \end{subfigure}
   \caption{BER and BLER under a Rayleigh channel.}
   \label{fig:Ray_ber}
 \end{figure}

CNN is employed in transmission with a large block size and the channel encoding is included.
We compare the end-to-end approach with a baseline method, where QAM is used as the modulation and the RSC of coderate $1/2$ are used as the coding.
In each block, $64$ information bits will be transmitted, thus the input size of the end-to-end approach is 64.
From Fig. \ref{fig:Ray_ber}, the end-to-end approach shows similar performance to the traditional methods in terms of BER and BLER, where in the baseline system, QAM is used as the modulation and the RSC is used as the coding.


\subsubsection{Frequency-Selective Fading Channel}
Under the frequency-selective channel, coded and uncoded end-to-end communication systems are developed with CNN and the OFDM system is used as the baseline.
There are 64 subcarriers in the OFDM system and the length cycle-prefix is set as 16 and 4 QAM is used for the modulation. In the coded system, the RSC coding is adopted.
In order to have a fair comparison, we set the block size of the end-to-end system as 64 bits and pad 16 zeros between every two blocks.

 \begin{figure}[ht]
  \begin{subfigure}[b]{0.5\linewidth}
    \centering
    \includegraphics[width=1\linewidth]{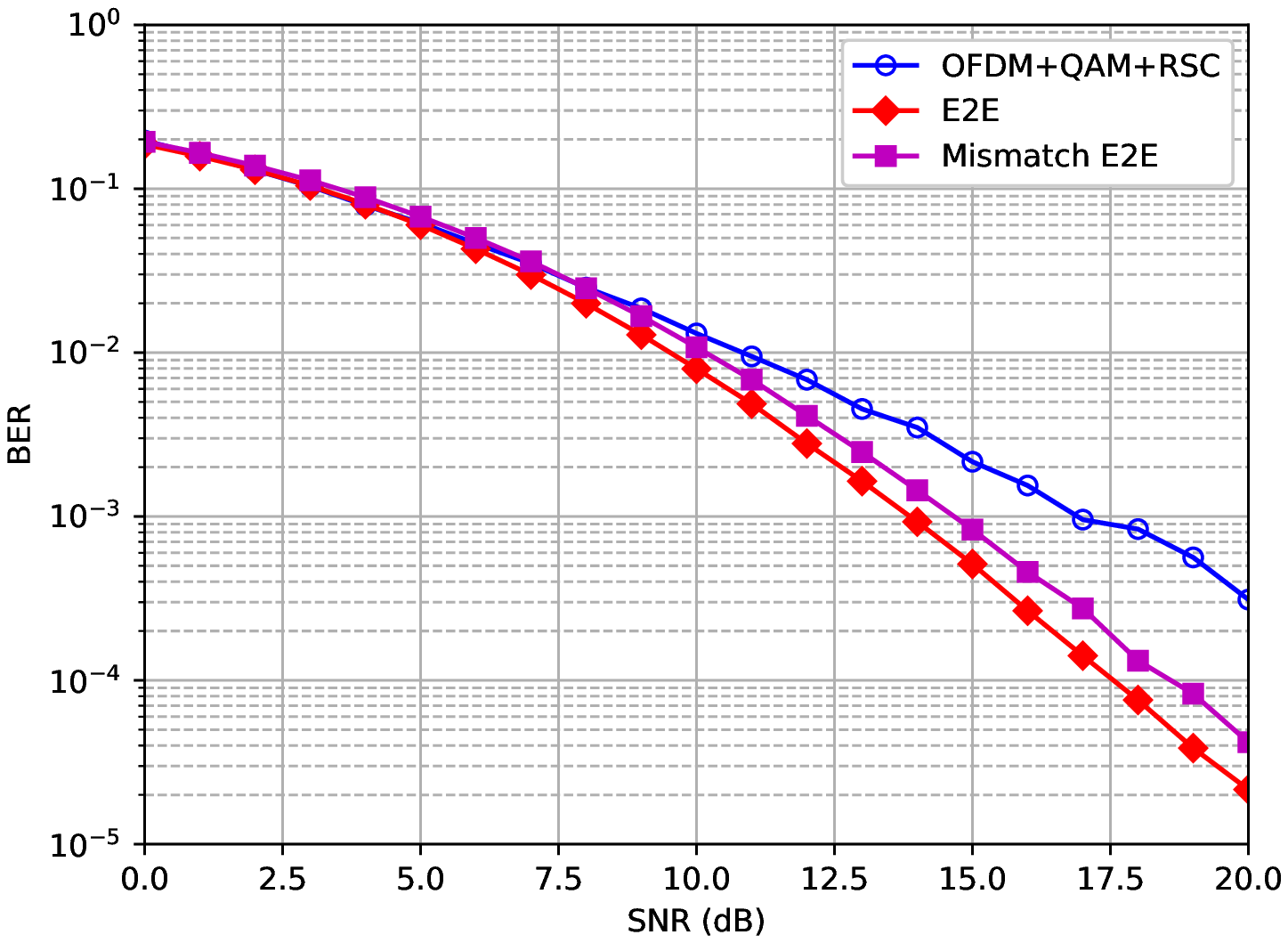}
  \end{subfigure}
  \begin{subfigure}[b]{0.5\linewidth}
    \centering
     \includegraphics[width=1\linewidth]{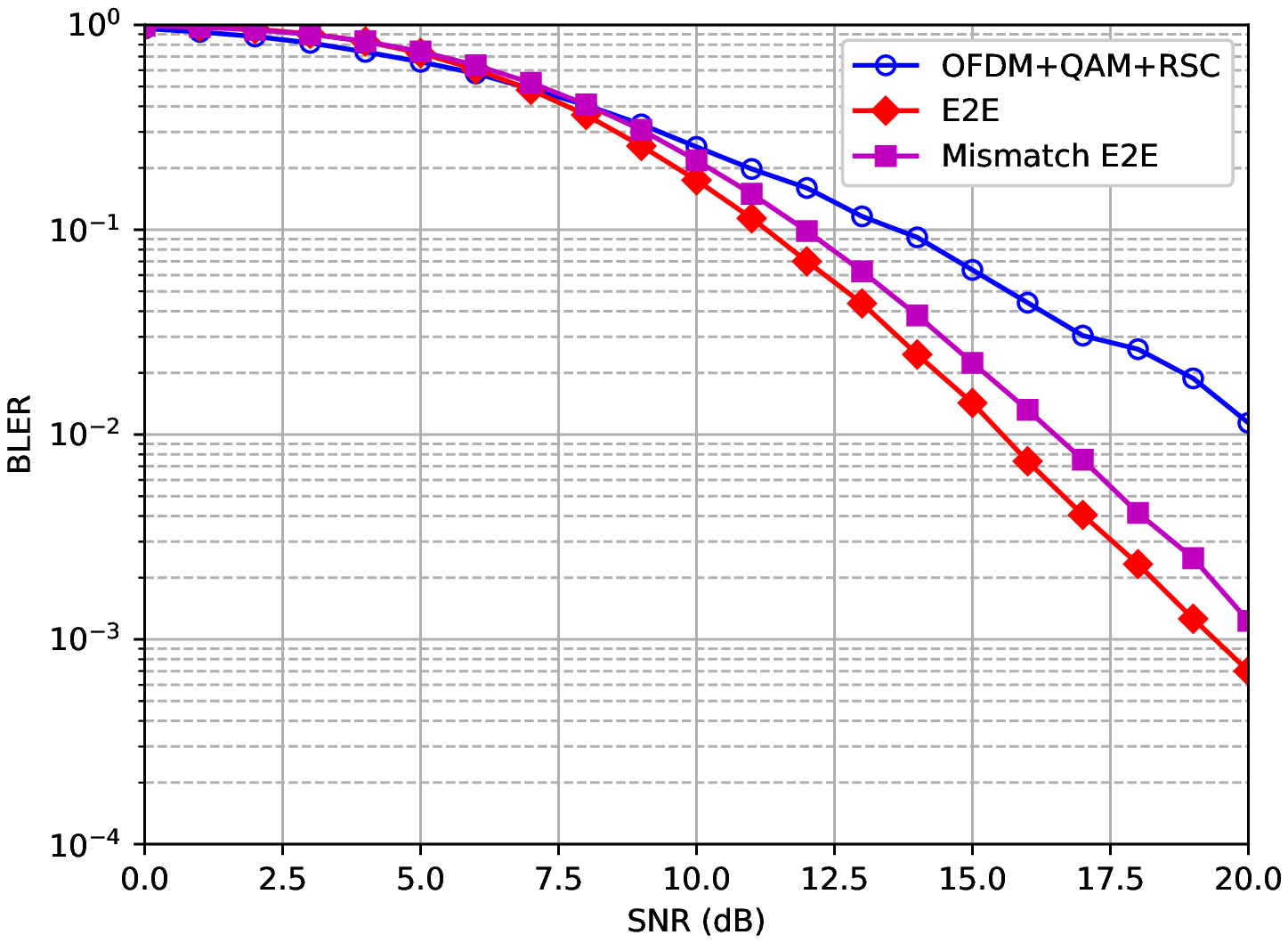}
   \end{subfigure}
   \caption{BER and BLER  under a frequency-selective multi-path channel.}
   \label{fig:OFDM_ber}
 \end{figure}


Fig.~\ref{fig:OFDM_ber} shows the performance of the propose end-to-end approach
The input size of end-to-end approach 64 bits.
From Fig.~\ref{fig:OFDM_ber},  the proposed end-to-end system significantly outperforms the OFDM system when the SNR is over 7 dB.
In addition, the mismatch of training and testing is considered. In the training stage, an exponential decay PDP, $\mathbb{E}|b_k|^2 = \frac{1}{2^k}$, is used to generate the channels while in the testing stage the equal strength PDP is used for evaluation. From the figure, even though the PDPs used in the training and testing are different to a large degree, the performance of the end-to-end model trained with mismatch is comparable to the model without mismatch, still much better than the OFDM system, which proves the robustness of the proposed method to the discrepancy between training and testing.

\section{Conclusions and Discussions} \label{sec:Conclusion}

In this article, we investigate the end-to-end learning of a communication system without prior information of the channel. We show that the conditional distribution of the channel can be modeled by a conditional GAN.
In addition, by adding the pilot information into the condition information, the conditional GAN can generate data corresponding to the specific instantaneous channel.

The end-to-end pipeline consists of DNNs for the transmitter, the channel GAN, and the receiver. By iteratively training these networks, the end-to-end loss can be optimized in a supervised way. The simulation results on the AWGN channles, Rayleigh fading channels, and frequency-selective channels confirm the effectiveness of the proposed method, by showing similar or better performance compared with the traditional approaches based on expert knowledge and channel models.
Our research opens a new door for building the pure data-driven communication systems.

One of the future directions is to test the proposed method in real data. As we have indicated in the introduction, in the real communication scenario, many imperfections will make the real channel difficult to express, which is very suitable for modeling these effects in a data-driven manner.

\bibliographystyle{IEEEbib}
\bibliography{strings,refs}

\end{document}